\def\year{2017}\relax
\documentclass[letterpaper]{article} %DO NOT CHANGE THIS

\usepackage{aaai18}  %Required
\usepackage{times}  %Required
\usepackage{helvet}  %Required
\usepackage{courier}  %Required
\usepackage{url}  %Required
\usepackage{graphicx}  %Required

\usepackage{epsfig}
\usepackage{amssymb}
\usepackage{amsmath}
\usepackage{amsfonts}

\usepackage{xspace}
\usepackage{todonotes}
\usepackage{hyphenat}
\usepackage{booktabs}
\usepackage{enumerate}
\usepackage{breqn}

\usepackage{pgfplots}
\usepackage{filecontents}

\newcommand{\limes}{\textsc{Limes}\xspace}
\newcommand{\mandolin}{\textsc{Mandolin}\xspace}

\begin{document}
\title{Mandolin: A Knowledge Discovery Framework for the Web of Data}

\author{
Tommaso Soru\\
University of Leipzig, Germany\\\texttt{tsoru@informatik.uni-leipzig.de}
\And
Diego Esteves\\
University of Bonn, Germany\\\texttt{esteves@cs.uni-bonn.de}
\AND
Edgard Marx\\
Leipzig University of Applied Sciences, Germany\\\texttt{edgard.marx@htwk-leipzig.de}
\And
Axel-Cyrille Ngonga Ngomo\\
University of Paderborn, Germany\\\texttt{axel.ngonga@upb.de}
}

\maketitle

\begin{abstract}
% Among the approaches for knowledge discovery in non-relational databases which have been proposed so far, many are based on Markov Logic Networks (MLNs). 
% However, the use of their original implementation on large datasets has been discouraged, as current frameworks suffer from a high computational complexity. 
Markov Logic Networks join probabilistic modeling with first-order logic and have been shown to integrate well with the Semantic Web foundations. 
While several approaches have been devised to tackle the subproblems of rule mining, grounding, and inference, no comprehensive workflow has been proposed so far. 
In this paper, we fill this gap by introducing a framework called \mandolin, which implements a workflow for knowledge discovery specifically on RDF datasets.
Our framework imports knowledge from referenced graphs, creates similarity relationships among similar literals, and relies on state-of-the-art techniques for rule mining, grounding, and inference computation. 
We show that our best configuration scales well and achieves at least comparable results with respect to other statistical-relational-learning algorithms on link prediction.
% We show that this additional information allows the discovery of links even between different knowledge bases. 
% Finally, we show that our approach is more scalable than other MLN frameworks
% with the added benefit of providing explanations for links in the form of human-readable rules.
\end{abstract}

\section{Introduction} \label{sec:intro}

The \emph{Linked Data cloud} has grown considerably since its inception.
To date, the total number of facts exceeds 130 billion, spread in over 2,500 available datasets.\footnote{Retrieved on February 15, 2017, from \url{http://lodstats.aksw.org/}.}
This massive quantity of data has thus become an object of interest for disciplines as diverse as Machine Learning~\cite{spohr2011machine,nikolov2012unsupervised,rowe2011predicting}, Evolutionary Algorithms~\cite{wang2006gaom,NGLY12}, Generative Models~\cite{bhattacharya2006latent}, and Statistical Relational Learning (SRL)~\cite{singla2006entity}.
%\todo[inline]{Citations needed}
One of the main objectives of the application of such algorithms is to address the fourth Linked Data principle, which
% This ten-year-old principle 
% preaches to the Semantic Web community 
states \textit{``include links to other URIs, so that they [the visitors] can discover more things''}~\cite{berners2006linked}.
Two years later, \cite{domingos2008just} proposed Markov Logic Networks (MLNs) -- a well-known approach to Knowledge Discovery in knowledge bases~\cite{richardson2006markov} -- to be a promising framework for the Semantic Web.
Bringing the power of probabilistic modeling to first-order logic, MLNs associate a weight to each formula (i.e., first-order logic rule) and are able to natively perform probabilistic inference.
Several tools based on MLNs have been designed so far~\cite{kok2009alchemy,niu2011tuffy,noessner2013rockit,bodart2014arthur}.
Yet, none of the existing MLN frameworks develops the entire pipeline from the generation of rules to the discovery of new relationships in a dataset.
Moreover, the size of the Web of Data represents today an enormous challenge for such learning algorithms, which often have to be re-engineered in order to scale to larger datasets.
In the last years, this problem has been tackled by proposing algorithms that benefit of massive parallelism.
Approximate results with some confidence degree have been preferred over exact ones, as they often require less computational power, yet leading to acceptable performances.

In this paper, we propose a new workflow for probabilistic knowledge discovery on RDF datasets and implement it in a framework called \mandolin, \underline{Ma}rkov Logic \underline{N}etworks for the \underline{D}iscovery \underline{o}f \underline{Lin}ks.
To the best of our knowledge, our framework is the first one to implement the entire pipeline for link prediction on RDF datasets.
% We then evaluate the goodness of the discovered rules by plugging to state-of-the-art techniques for the subtasks of grounding and inference computation.
Making use of RDFS/OWL semantics, \mandolin can (i) import knowledge from referenced graphs, (ii) compute the forward chaining, and (iii) create similarity relationships among similar literals.
We show that this additional information allows the discovery of links even between different knowledge bases.
% Finally, we show that all the components of our framework scale well. 
We evaluate the framework on two benchmark datasets for link prediction and show that it can achieve comparable results w.r.t. other SRL algorithms and outperform them on two accuracy indices.
%\todo[inline]{The property similarities are not mentioned.}

% This paper is structured as follows.
% The next section presents the related work.
% We then start with some preliminaries and we describe the workflow in details.
% After showing the experiments and their discussion, we conclude with a list of future works. 
% We then start with some preliminaries in Section~\ref{sec:prel}; afterwards, we describe the workflow in details in Section~\ref{sec:mandolin}.
% Section~\ref{sec:exp} shows the experiments. %, which are discussed in Section~\ref{sec:discussion}.
% Finally, we conclude with a list of future works. 

\section{Related Work} \label{sec:related}

Machine-learning techniques have been successfully applied to ontology and instance matching, where the aim is to match classes, properties, and instances belonging to different ontologies or knowledge bases~\cite{NGO+11,NGLY12,DBLP:conf/semweb/2015om}.
Also, evolutionary algorithms have been used to the same scope~\cite{martinez2008optimizing}.
For instance, genetic programming has shown to find good link specifications (i.e., similarity-based decision trees) in both a semi-supervised and unsupervised fashion~\cite{NGLY12}.
Generative models are statistical approaches which do not belong to the ML and SRL branches.
Latent Dirichlet allocation is an example of application to entity resolution~\cite{bhattacharya2006latent} and topic modeling~\cite{ESWC_Tapioca2016}.

SRL techniques such as Markov-logic~\cite{richardson2006markov} and tensor-factorization models~\cite{nickel2014reducing} have been proposed for link prediction and triple classification; the formers have also been applied on problems like entity resolution~\cite{singla2006entity}.
% \todo[inline]{SRL methods (MLN frameworks, factorizers, KGE algorithms)}
Among the frameworks which operate on MLNs, we can mention NetKit-SRL~\cite{macskassy2005netkit}, Alchemy~\cite{kok2009alchemy}, Tuffy~\cite{niu2011tuffy}, ArThUR~\cite{bodart2014arthur}, and RockIt~\cite{noessner2013rockit}.
Several approaches which rely on translations have been devised to perform link prediction via generation of embeddings~\cite{TransE/bordes2013translating,TransR/lin2015learning,TransRrules/wang2015knowledge,TransG/xiao2015transg}.
The Google Knowledge Vault is a huge structured knowledge repository backed by a probabilistic inference system (i.e., ER-MLP) that computes calibrated probabilities of fact correctness~\cite{dong2014knowledge}.

This work is also related to link prediction in social networks~\cite{liben2007link,scellato2011exploiting}.
Being social networks the representation of social interactions, they can be seen as RDF graphs having only one property.
Recently, approaches such as DeepWalk~\cite{perozzi2014deepwalk} and node2vec~\cite{grovernode2vec} showed impressive scalability to large graphs.

The link discovery frameworks Silk~\cite{volz2009silk} and \limes~\cite{NGAU11} present a variety of methods for the discovery of links among different knowledge bases~\cite{jentzsch2010silk,NGLY12,sherifDPSO}.
As presented in the next section, instance matching is a sub-problem of link discovery where the sought property is an equivalence linking instances.
Most instance matching tools~\cite{li2009rimom,jimenez2011logmap} take or have taken part in the \emph{Ontology Matching Evaluation Initiative} (OAEI).

\begin{figure*}[t]
    \centering
    \includegraphics[width=0.5\textwidth]{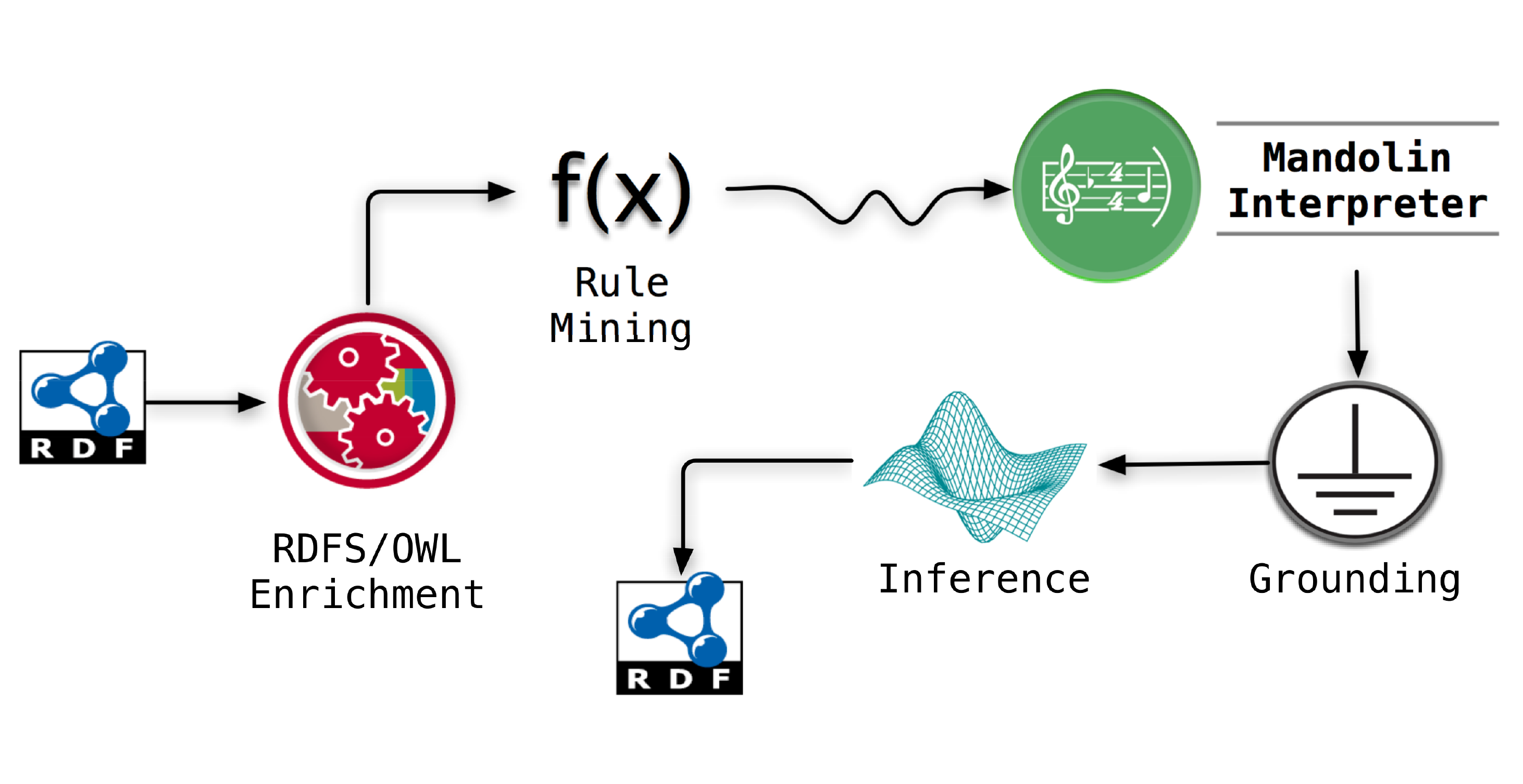}
    \caption{The knowledge discovery workflow as implemented in \mandolin.}
    \label{fig:overview}
\end{figure*}

\section{Preliminaries} \label{sec:prel}

First-order knowledge bases are composed of statements and formulas expressed in first-order logic~\cite{Genesereth:1987:LFA:31838}.
In probabilistic knowledge bases, every statement (i.e., edge) has a weight associated with it~\cite{wuthrich1995probabilistic}.
The weighting function can be represented as $\omega : E \rightarrow [0,1]$.
This means that a relationship might exist within some confidence or probability degree.
The current Semantic Web vision does not foresee weights for a given triple.
% In the current Semantic Web vision, any existing relationship (i.e., triple) is an edge having weight 1. \todo[inline]{Can be regarded as having a weight of 1? Formally, the SemWeb does not foresee weights}
However, a probabilistic interpretation of RDF graphs with weights partly lower than 1 has shown to be able to help to solve many problems, such as instance matching, question answering, and class expression learning~\cite{leitao2007structure,SHE+14,Buehmann2014}.

% \subsection{Markov Logic Networks} \label{sec:mln}

As mentioned in the introduction, MLNs join first-order logic with a probabilistic model by assigning a weight to each formula.
Formally, a \emph{Markov Logic Network} can be described as a set $(F_i, w_i)$ of pairs of formulas $F_i$, expressed in first-order logic, and their corresponding weights $w_i \in \mathbb{R}$.
The weight $w_i$ associated with each formula $F_i$ softens the crisp behavior of Boolean logic as follows.
Along with a set of constants $C$, an MLN can be viewed as a template for building a Markov Network.
Given $C$, a so-called \emph{Ground Markov Network} is thus constructed, leaving to each grounding the same weight as its respective formula.
% The distribution of all possible worlds is then defined as:
% \begin{equation} \label{eqn:allworlds}
% P(X=x) = \frac{1}{Z} \exp{\left( \sum_i w_i n_i(x) \right)}
% \end{equation}
% where $n_i(x)$ is the number of true groundings of $F_i$ in $x$~\cite{richardson2006markov}.
% Please note that, despite (\ref{eqn:distr2}) and (\ref{eqn:allworlds}) might look similar, the two summations iterate on different entities, i.e. cliques and MLN formulas, respectively.
% The nodes may assume a value of $1$ if the ground predicate is true or $0$ if false.
% Similarly, if the formula is true, its feature assumes the value of $1$ and $0$ otherwise.
In a ground network, each ground node corresponds to a statement.
Once such network is built, it is possible to compute a probability value $\in [0,1]$ for each statement~\cite{richardson2006markov}.

% \subsection{Research questions}

% With this work, we aim at answering to the following research questions:

% \begin{enumerate}[ {(Q}1{)} ]
%     \item Can MLNs outperform other SRL techniques on link prediction?
%     \item Can we optimize the trade-off between their computation needs and accuracy?
%     \item Can MLNs scale to large datasets?
%     \item Can we join RDF semantics and Markov rules to improve the predictions?
%     \item Can we use MLNs as an approach to RDF instance matching?
% \end{enumerate}

% In the following section, we present our proposed resolution to the research questions above, whereas in Section~\ref{sec:exp}, we will show the results we obtained.

\section{The Workflow} \label{sec:mandolin}

Our workflow comprises five modules: \emph{RDFS/OWL enrichment}, \emph{Rule mining}, \emph{Interpretation}, \emph{Grounding}, and \emph{Inference}.
As can be seen in Figure~\ref{fig:overview}, the modules are aligned in a sequential manner.
Taking a union of RDF graphs $G=\bigcup_i G_i$ as input, the process ends with the generation of an enriched graph $G'=(V',E')=lp(G)$ where $lp$ is the link prediction algorithm modeled as function.

\subsection{RDFS/OWL enrichment} \label{sec:enrichment}

The \emph{RDFS/OWL enrichment} module activates optionally and features three different operations: \emph{Similarity join}, \emph{Ontology import}, and \emph{Forward chaining}.
Its function is to add a layer of relationships to the input graph $G$.

\subsubsection{Similarity join.}

A node in an RDF graph may represent either a URI, a literal, or a blank node.
% \todo[inline]{properties cannot be blank nodes}
While URI or a blank node have no restrictions w.r.t. their end in the triple (i.e., they can both be subjects or objects), a literal can only be put as an object (i.e., have only incoming edges).
Literals can be of different datatype (e.g., strings, integers, floats).
In order to generate the similarity relationships, we first collect all literals in the graph into as many buckets as there are datatype properties.
We chose to use the Jaccard similarity on \emph{q-grams}~\cite{Gravano:2001:ASJ:645927.672200} to compare strings.
To tackle the quadratic time complexity for the extraction of similar candidate pairs, we apply a positional filtering on prefixes and suffixes~\cite{Xiao:2008:ESJ:1367497.1367516} as implemented in the \limes framework~\cite{NGON12} within a similarity threshold $\theta$.
Once the candidate pairs (i.e., datatype values) are extracted, we create a new similarity property for each datatype property and for $\theta=0.1,...,1.0$ to connect the respective subjects (e.g., \texttt{:foaf\_name\_0.6} to link two persons having names with similarity greater than $0.6$).
% \todo[inline]{The sentence above is unclear.}
% For example, let us set $\theta=0.6$ and consider the following triples:
% {\footnotesize\begin{verbatim}
%  :New_York_City :isIn :New_York
%  :New_York :isIn :USA
%  :New_York_City foaf:name "New York City"@en
%  :New_York foaf:name "New York"@en
%  :USA foaf:name "USA"@en
% \end{verbatim}}
% we first collect strings \texttt{New York City}, \texttt{New York}, and \texttt{USA}.
% After the filtering, the only extracted pair for property \texttt{foaf:name} is $($\texttt{New York City}$,~$\texttt{New York}$)$, having a Jaccard similarity of $8/13=0.61$.
% We generate a new URI featuring the \emph{SHA1} hash function of the original property URI and the threshold value for a new property, such as:
% {\footnotesize\begin{verbatim}
% http://mandolin.aksw.org/similarity/
%     d0c70c5ef3a2cd1e38e266bcf5e2d607e4bbd47f/0.6
% \end{verbatim}}
% which is then used, for each extracted pair, to connect the two subjects linked to them via \texttt{foaf:name}, i.e. \texttt{:New\_York\_City} and \texttt{:New\_York}.
Intuitively, these similarity properties form a hierarchy where properties with a higher threshold are sub-properties of the ones with lower threshold.
% As large multi-domain datasets such as DBpedia contains $n=5,729$ properties, we estimate the probability of a hash collision as $p=\frac{n(n-1)}{2^{b+1}} \approx 10^{-41}$.

The procedure above is repeated for each datatype property.
Numerical and time values, are sorted by value and linked via a similarity predicate whenever their difference is less than the threshold $\theta$.
The rationale behind the use of similarity joins is that (i) they can foster the discovery of equivalence relationships and (ii) similarity properties can be included in inference rules. 

\subsubsection{Ontology import and Forward chaining.}

RDF datasets on the Web are published so that their content can be accessible from everywhere.
The vision of the Semantic Web expects URIs to be referenced from different knowledge bases.
In any knowledge-representation application, one option to process the semantics associated with a URI is to import the ontology (or the available RDF data) which defines such entity.
To accomplish this, \mandolin dereferences external URIs, imports the data into its graph $G$, and performs forward chaining (i.e., semantic closure) on the whole graph.
% For instance, importing the declaration of \texttt{foaf:name}, we notice it is a sub-property of \texttt{rdfs:label}.
% Which means that, after performing the forward chaining, all \texttt{foaf:name} relationships exist also as \texttt{rdfs:label}s.
This additional information can be useful for the Markov logic, since it fosters connectivity on $G$.

\subsection{Rule mining and Interpretation}

The mining of rules in a knowledge base is not a task strictly related to MLN systems.
Instead, the set of MLN rules is usually given as input to the MLN system.
\mandolin integrates the rule mining phase in the workflow exploiting a state-of-the-art algorithm.

The rule mining module takes an RDF graph as input and yields rules expressed as first-order Horn clauses.
A Horn clause is a logic clause having at most one positive literal if written in the disjunctive normal form (DNF).
Any DNF clause $\neg a(x,y) \vee c(x,y)$ can be rewritten as $a(x,y) \Rightarrow c(x,y)$, thus featuring an implication.
The part that remains left of the implication is called \emph{body}, whereas the right one is called \emph{head}.
In \mandolin, a rule can have a body size of 1 or 2, belonging to one of 6 different classes:
$a(x,y) \Longrightarrow c(x,y)$, $a(y,x) \Longrightarrow c(x,y)$, $a(z,x) \wedge b(z,y) \Longrightarrow c(x,y)$, $a(x,z) \wedge b(z,y) \Longrightarrow c(x,y)$, $a(z,x) \wedge b(y,z) \Longrightarrow c(x,y)$, and $a(x,z) \wedge b(y,z) \Longrightarrow c(x,y)$.
% In \mandolin, a rule can belong to one of the following classes:
% \begin{enumerate}
%     \item $a(x,y) \Longrightarrow c(x,y)$
%     \item $a(y,x) \Longrightarrow c(x,y)$
%     \item $a(z,x) \wedge b(z,y) \Longrightarrow c(x,y)$
%     \item $a(x,z) \wedge b(z,y) \Longrightarrow c(x,y)$
%     \item $a(z,x) \wedge b(y,z) \Longrightarrow c(x,y)$
%     \item $a(x,z) \wedge b(y,z) \Longrightarrow c(x,y)$
% \end{enumerate}
% where, for our notation, a statement $a(x,y)$ is an edge $e=(x,y) \in E$ such that $l(e)=\tt{a}$.
% Note that the universal quantifiers have been omitted since the rules are declared in a non-propositional way.
Intuitively, considering only a subset of Horn clauses decreases expressivity but also the search space.
In large-scale knowledge bases, this strategy is preferred since it allows to scale.
% 
% % #x <==> \{x : X \in \mathcal{K}\}
% Rules in knowledge bases can be ranked using several indices.
% The \emph{support} of a rule is defined as the number of correct predictions in the data.
% For instance, the support ($\sigma$) for rules of class 3 is so defined: % classes 1 and 3 respectively is so defined:
% % \begin{multline}
% % \sigma(a(x,y) \Longrightarrow b(x,y)) := |(x,y) \in E| : a(x,y) \wedge b(x,y) 
% % \end{multline}
% \begin{multline}
% \sigma(a(z,x) \wedge b(z,y) \Longrightarrow c(x,y)) := \\
% |\{(x,y) \in E : \exists z : a(z,x) \wedge b(z,y) \wedge c(x,y)\}|
% \end{multline}
% However, as these values are absolute, a more proper measure was proposed~\cite{galarraga2015fast} to maintain independence from the size of the graph.
% The \emph{head coverage} ($\eta$) of a rule $F \in \mathcal{F}$ is a normalized version of support and is defined as follows.
% \begin{equation} \label{eqn:hc}
% \eta(F) := \frac{\sigma(F)}{|\{e \in E : l(e)=\tt{c}\}|}
% \end{equation}
% Finally, a measure of the confidence ($\kappa$) of a rule $F$ is introduced.
% This index is also referred to as \emph{Partial Completeness Assumption (PCA) confidence}~\cite{galarraga2015fast}:
% \begin{equation} \label{eqn:pca}
% \kappa(F) := 
% \frac{\sigma(F)}{|\{e=(x,y) \in E : \exists z_1, \ldots, z_m, y' : l(e)=c \wedge (x,y') \in E\}|}
% \end{equation}
% (\ref{eqn:hc}) and (\ref{eqn:pca}) play an important role in the rule mining phase, as their values indicate whether or not a rule has to be pruned from the results.
For the search of rules in the graph, we rely on the \emph{AMIE+} algorithm described in~\cite{galarraga2015fast}.

In the interpretation module, the set of rules returned by the rule miner is collected, filtered, and translated for the next phase, i.e. the grounding.
At the end of the mining phase, we perform a selection of rules based on their head coverage, i.e. $F'=\{F \in \mathcal{F} : \eta(F) \geq \bar{\eta}\}$.
We preferred to use PCA confidence over head coverage because previous literature showed its greater effectiveness~\cite{dong2014knowledge,galarraga2015fast}.

% \subsection{Interpretation}

% In this module, the set of rules outputted by the rule miner are collected, filtered, and translated for the next phase, i.e. the grounding.
% At the end of the mining phase, we perform a selection of rules based on their head coverage, i.e. $F'=\{F \in \mathcal{F} : \eta(F) \geq \bar{\eta}\}$.
% In most MLN frameworks, weights associated to rules are provided manually~\cite{jain2010soft,noessner2013rockit,bodart2014arthur}; some systems can learn them from data~\cite{kok2009alchemy,niu2011tuffy}.
% However, as in our framework the rules come with their respective measures, a weight learning phase is not needed.
% % weight := k * PCA confidence (k=100)
% We thus assign a weight to a rule using the following formula:
% \begin{equation}
% w(F) := k * \kappa(F)
% \end{equation}
% with $k=100$.
% We preferred to use PCA confidence over head coverage because previous literature showed its greater effectiveness~\cite{dong2014knowledge,galarraga2015fast}.

\subsection{Grounding}

Grounding is the phase where the ground Markov network (factor graph) is built starting from the graph and a set of MLN rules.
A factor graph is a graph consisting of two types of nodes: factors and variables where the factors connect all the variables in their scope. 
Given a set of factors $\phi=\{\phi_1, \ldots, \phi_N\}$, $\phi_i$ is a function over a random vector $X_i$ which indicates the value of a variable in a formula.
As the computational complexity for grounding is NP-complete, the problem of scalability has been addressed by using relational databases.
However, frameworks such as \emph{Tuffy} or \emph{Alchemy} showed they are not able to scale, even in datasets with a few thousand statements~\cite{chen2013web}.
Tuffy, for example, stores the ground network data into a DBMS loaded on a RAM-disk for best performances~\cite{niu2011tuffy};
however, growing exponentially, the RAM cannot contain the ground network data, resulting in the program going out of memory.
For this reason, in the \mandolin module for grounding, we integrated \emph{ProbKB}, a state-of-the-art algorithm for the computation of factors.
The main strength of this approach is the exploitation of the simple structure of Horn clauses~\cite{chen2014knowledge}, differently from other frameworks where any first-order logic formula is allowed.
It consists of a two-step method, i.e. (1) new statements are inferred until convergence and (2) the factor network is built.
Each statement is read in-memory at most 3 times; differently from Tuffy, where it is read every time it appears in the knowledge base~\cite{chen2013web}.

\subsection{Inference}

MAP Inference in Markov networks is a P\#-complete problem~\cite{Roth:1996:HAR:227773.227790}.
However, the final probability values can be approximated using techniques such as Gibbs sampling -- which showed to perform best~\cite{noessner2013rockit} -- and belief propagation~\cite{richardson2006markov}.
%In our workflow, we employ the use of Gibbs sampling. for the computation of the set of links $P$ defined in~(\ref{eqn:aim_lp}).
% We employ the use of Gibbs sampling for the computation of the set of links $P$ defined in~(\ref{eqn:aim_lp}) and~(\ref{eqn:aim_di}).
% Typically, the number of iterations for the Gibbs sampler is $\gamma=100*|E|$~\cite{noessner2013rockit}.
% Due to scalability reasons, we approximate the probability values by limiting the number of iterations.
% \todo[inline]{Do we really need the two sentences above?}

Every statement $a(x,y)$ is associated with a node in the factor graph.
Therefore, its probability is proportional to the product of all potential functions $\phi_k(x_{\{k\}})$ applied to the state of the cliques touching that node.
Once we compute the probabilities of all sampled candidates, we normalize them so that the minimum and maximum values are $0$ and $1$ respectively. %instead of dividing the product by the partition function constant $Z$ (see Section~\ref{sec:mln}),
The final set $P$ of predicted links is then defined as those statements whose probability is greater than a threshold $\tau \in [0,1]$.
Typically, the number of iterations for the Gibbs sampler is $\gamma=100*|E|$~\cite{noessner2013rockit}.

% \subsection{Incremental learning}

% Another peculiarity of our framework is the incremental learning setting.
% After the set of links $P$ has been discovered, \mandolin can optionally run again on the enriched graph $G'$ to yield a more enriched graph $G''$.
% We use a validation set to let the algorithm estimate the performances and decide whether to stop.
% In case the performance of the last iteration is the highest found so far, another iteration is carried out. 

% \subsection{Implementation} \label{sec:impl}

% The \mandolin framework was mainly developed in \emph{Java 1.8} and its source code is available online\footnote{\url{https://github.com/AKSW/Mandolin}} along with all datasets used for the evaluation.
% % The \mandolin framework was mainly developed in \emph{Java 1.8} and its source code will be made available online.\footnote{Binaries and data can be temporarily downloaded at \url{http://bit.ly/2lCn5eV}.} along with all datasets used for the evaluation.
% Libraries and research projects involved are \emph{Apache Jena}, \emph{Pellet}, \emph{AMIE+}, \emph{ProbKB}, and the \emph{Gurobi} optimizer.
% To implement the Gibbs sampling, we chose to use \emph{RockIt}~\cite{noessner2013rockit} instead of \emph{GraphLab}~\cite{low2010graphlab}, since the latter project had just been rebranded and its code privatized.
% We rely on a \emph{PostgreSQL} database for the storage of the networks.
% How the input graph and MLN rules are stored and the factor graph computed via a \emph{Pl/PgSQL} script is described in~\cite{chen2014knowledge}.

\section{Experiments} \label{sec:exp}
% \subsection{Evaluation setup}

Any directed labeled graph can be easily transformed into an RDF graph by simply creating a namespace and prepending it to entities and properties in statements.
We thus created an RDF version of a benchmark for knowledge discovery used in \cite{TransE/bordes2013translating,TransR/lin2015learning,TransRrules/wang2015knowledge,TransG/xiao2015transg,nickel2015holographic}.
The benchmark consists of two datasets: \emph{WN18}, built upon the WordNet glossary, and \emph{FB15k}, a subset of the Freebase collaborative knowledge base.
Using these datasets, we evaluated \mandolin on link prediction.
% Moreover, we used two datasets from the \emph{Instance Matching} track of the OAEI benchmark to evaluate our approach on data integration.\footnote{\url{http://oaei.ontologymatching.org}}
% Finally, we employed two large-scale datasets to evaluate the scalability of our approach.
Finally, we employed the DBLP-ACM~\cite{DBLP:journals/pvldb/KopckeTR10} dataset to test cross-dataset linking, as well as the large-scale dataset DBpedia 3.8 to evaluate the scalability of our approach.
All experiments were carried out on a 64-core server with 256 GB RAM equipped with Ubuntu 16.04.
The \mandolin framework %\footnote{DOI: \url{https://doi.org/10.6084/m9.figshare.5007998}.}
was mainly developed in \emph{Java 1.8} and its source code is available online\footnote{\url{https://github.com/AKSW/Mandolin}} along with all datasets used for the evaluation.

\begin{table}[h]
\centering
\begin{tabular}{lrrr}
\toprule
\textbf{Dataset} & \textbf{\# triples} & \textbf{\# nodes} & \textbf{\# prop.} \\
\midrule
{WN18}     & 146,442     & 40,943    & 18 \\
{FB15k}   & 533,144     & 14,951    & 1,345 \\
{DBLP--ACM} & 20,759 & 5,003 & 34 \\
{DBpedia 3.8} & 11,024,066 & $\sim$2,200,000 & 650 \\
% {DBpedia} & 397,831,457 & 5,174,547 & 63,764 \\
% {Wikidata} &  &  &  \\
\bottomrule
\end{tabular}
\caption{Datasets used in the evaluation.}
\label{tab:datasets}
\end{table}

\begin{table*}[htb]\small
\centering
\caption{Results for link prediction on the WordNet (WN18) and Freebase (FB15k) datasets.}\label{tab:lp2}
\begin{tabular}{lcccccccc}
\toprule
             & \multicolumn{4}{c}{\textbf{WN18}}                              & \multicolumn{4}{c}{\textbf{FB15k}}                             \\ \cmidrule(lr){2-5} \cmidrule(lr){6-9} 
             & MRR            & Hits@1        & Hits@3        & Hits@10       & MRR            & Hits@1        & Hits@3        & Hits@10       \\ \midrule
\textsc{TransE}       & 0.495          & 11.3          & 88.8          & 94.3          & 0.463          & 29.7          & 57.8          & \textbf{74.9} \\
\textsc{TransR}       & 0.605          & 33.5          & 87.6          & 94.0          & 0.346          & 21.8          & 40.4          & 58.2          \\
\textsc{ER-MLP}       & 0.712          & 62.6          & 77.5          & 86.3          & 0.288          & 17.3          & 31.7          & 50.1          \\
\textsc{RESCAL}       & 0.890          & 84.2          & 90.4          & 92.8          & 0.354          & 23.5          & 40.9          & 58.7          \\
\textsc{HolE}         & \textbf{0.938} & \textbf{93.0} & \textbf{94.5} & 94.9          & \textbf{0.524} & 40.2          & \textbf{61.3} & 73.9          \\ \midrule
\mandolin & 0.892          & 89.2          & 94.3          & \textbf{96.0} & 0.404          & \textbf{40.4} & 48.4          & 52.6 \\
% \mandolin (1) & 0.892          & 89.2          & 94.3          & \textbf{96.0} & 0.307          & 30.7          & 36.7          & 39.8 \\
% \mandolin (2) & 0.780          & 78.0          & 82.2          & 84.0          & 0.404          & \textbf{40.4} & 48.4          & 52.6  \\ 
\bottomrule
\end{tabular}
\end{table*}

% \subsection{Link prediction evaluation}

We evaluated the link prediction task on two measures, \emph{Mean Reciprocal Rank} (MRR) and \emph{Hits@k}.
The benchmark datasets are divided into training, validation, and test sets.
We used the training set to build the models and the validation set to find the hyperparameters, which are introduced later.
Afterwards, we used the union of the training and validation sets to build the final model.
For each test triple $(s,p,o)$, we generated as many corrupted triples $(s,p,\tilde{o})$ as there are nodes in the graph such that $o \neq \tilde{o} \in V$.
We computed the probability for all these triples when this value was available; when not, we assumed it $0$.
Then, we ranked the triples in descending order and checked the position of $(s,p,o)$ in the rank.
% The MRR is thus the reciprocal rank, averaged to all test triples:
% \begin{equation}
% MRR = \frac{1}{Q} \sum_{i=1}^{|Q|} \frac{1}{rank_i}
% \end{equation}
% MRR has been preferred over mean rank because it is less sensitive to outliers~\cite{nickel2015holographic}.
The Hits@k index is the ratio (\%) of test triples that have been ranked among the top-$k$.
We computed Hits@1,3,10 with a filtered setting, i.e. all corrupted triples ranked above $(s,p,o)$ which are present in the training sets were discarded before computing the rank.

The results for link prediction on the \emph{WN18-FB15k} benchmark are shown in Table~\ref{tab:lp2}.
We compare \mandolin with other SRL techniques based on embeddings and tensor factorization.
On \emph{WN18}, we overperform all other approaches w.r.t. the Hits@10 index (96.0\%).
However, \textsc{HolE}~\cite{nickel2015holographic} recorded the highest performance on MRR and Hits@1; the two approaches achieved almost the same value on Hits@3.
% Here, \mandolin recorded its own highest performance after 1 iteration of incremental learning; the second iteration saw a drop of $-8\%$ in all measures.
% On the Freebase dataset, three different approaches hold the highest values.
% \textsc{HolE} performed best on MRR and Hits@3, whereas \mandolin on Hits@1, and \textsc{TransE} on Hits@10.
% On this dataset, our incremental learning setting showed its effectiveness, yielding a $+12\%$ Hits@10 from the first to the second iteration, which recorded the highest local value. Examples of rules learned can be found at the project website.
% 
Since the two datasets above contain no datatype values and no statements using the RDF schema\footnote{\url{https://www.w3.org/TR/rdf-schema/}}, we did not activate the RDF-specific settings introduced in the previous section.

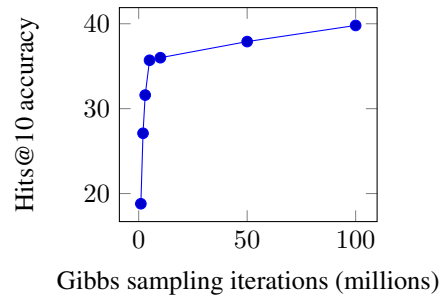
\begin{figure}[htp]
    \centering
    \begin{tikzpicture}
    \begin{axis}[
    	scale=0.50,
    	legend pos=outer north east,
    	xlabel=Gibbs sampling iterations (millions),
    	ylabel=Hits@10 accuracy
    ]
    % \addplot table [x=gibbs, y=hits1, col sep=tab] {mln.csv};
    % \addplot table [x=gibbs, y=hits3, col sep=tab] {mln.csv};
    \addplot table [x=samplings, y=hits10, col sep=tab] {mln.csv};
%     \legend{
% % 			hits1,
% % 			hits3,
% 			hits10,}
    \end{axis}
    \end{tikzpicture}

    \caption{Hits@10 on \emph{FB15k} with $\bar{\eta}=0.9$.}
    \label{fig:sampliter}
\end{figure}

Our framework depends on the following hyperparameters:
\begin{itemize}
    \item \emph{minimum head coverage} ($\bar{\eta}$), used to filter rules;
    \item \emph{Gibbs sampling iterations} ($\gamma$).
\end{itemize}
To compute the optimal configuration on the trade-off between computational needs and overall performances, we performed further experiments on the link prediction benchmark.
We investigated the relationship between number of Gibbs sampling iterations, runtime, and accuracy by running our approach using the following values: $\gamma=\{1,2,3,5,10,50,100\} \cdot 10^6$.
The runtime is, excluding an overhead, linear w.r.t. the number of iterations.
% Our findings showed that the runtime is, excluding an overhead, linear w.r.t. the number of iterations.
As can be seen in Figure~\ref{fig:sampliter}, the Hits@10 index tends to stabilize at around $\gamma=5 \cdot 10^6$, however higher accuracy can be found by increasing this value.
% The Hits@10 index seems to stabilize at around $\gamma=5M$, however higher accuracy can be found by increasing this value.
% We also performed tests on large-scale datasets such as DBpedia and Wikidata.
% In both cases, \mandolin was able to terminate the computation after at most $23.6$ hours, showing that our workflow can scale.

% \subsection{Scalability and instance matching}
% \subsection{Large-scale datasets}

We performed tests on the DBLP--ACM dataset for the discovery of equivalence relationships and on the large-scale dataset DBpedia\footnote{Version 3.8 from \url{http://www.mpi-inf.mpg.de/departments/ontologies/projects/amie/}.}.
% \todo[inline]{Talk about the tests vs Alchemy and its inefficient rule generation.}
We compared our approach with other MLN frameworks, i.e. NetKit-SRL, ProbCog, Alchemy, and Tuffy.
As these frameworks can learn rule weights but not rules themselves, we fed them with the rules found by our rule miner.
We set $\bar{\eta}=0.9$ and $\gamma=10^7$.
The results (see Table~\ref{tab:scala}) showed that, in all cases, \mandolin is the only framework that was able to terminate the computation.
In the DBLP--ACM dataset, we were able to discover equivalence links among articles and authors that had not been linked in the original datasets.
After dividing the mapping $M$ into two folds ($90\%-10\%$), we used the larger as training set.
We were able to predict $71.0\%$ of the correct \texttt{owl:sameAs} links in the remaining test set.
% As DBpedia and Wikidata are not manually generated, i.e. they cannot be considered gold standards, we could not measure any accuracy, however Table~\ref{tab:scala} shows that our system can deal even with such large datasets.
% \todo[inline]{Rephrase ``we could not measure any accuracy''.}

\begin{table}[ht]\small
    \centering
    \caption{Runtime, number of rules after the filtering, and number of predicted links for $\bar{\eta}=0.9$ and $\gamma=10^7$.}
    \begin{tabular}{l ccc}
        \toprule
        \textbf{Dataset} & \textbf{Runtime (s)} & ($|\mathcal{F}'|$) & ($|P|$) \\ \midrule
        DBLP--ACM & 2,460 & 1,500 & 4,730 \\
        DBpedia & 85,334 & 1,500 & 179,201 \\
        % Wikidata & 46,444 & 1,500 & 83,599 \\
        \bottomrule
    \end{tabular}
    \label{tab:scala}
\end{table}

% \subsection{Instance matching}

% \todo[inline]{Introduce briefly.}
% In the DBLP--ACM dataset, we were able to discover equivalence links among articles and authors that had not been linked in the original datasets.
% After dividing the mapping $M$ into two folds ($90\%-10\%$), we used the larger as training set.
% We were able to predict $71.0\%$ of the correct \texttt{owl:sameAs} links in the remaining test set. DBpedia and Wikidata do not provide a ground truth but we could show that we scale well even on such large datasets (see Table \ref{tab:scala}).

% \begin{table}[ht]
%     \centering
%     \caption{Runtime, number of rules after the filtering, and number of predicted links on the larger datasets are shown.}
%     \begin{tabular}{l ccc}
%         \toprule
%         \textbf{Dataset} & \textbf{Runtime (s)} & ($|\mathcal{F}'|$) & ($|P|$) \\ \midrule
%         DBLP--ACM & 2,460 & 1,500 & 4,730 \\
%         DBpedia & 85,334 & 1,500 & 179,201 \\
%         Wikidata & 46,444 & 1,500 & 83,599 \\
%         \bottomrule
%     \end{tabular}
%     \label{tab:scala}
% \end{table}

\section{Discussion} \label{sec:discussion}
We have witnessed a different behavior of our algorithm when evaluated on the two datasets for link prediction.
% In particular, the incremental learning setting showed to be beneficial only on the Freebase dataset.
This might be explained by the different structure of the graphs:
Relying on first-order Horn clauses, new relationships can only be discovered if they belong to a 3-vertex clique where the other two edges are already in the graph.
Therefore, rule-based algorithms might need one more step, such as longer body in rules or more iterations, to discover them on a less connected graph such as FB15k.
A more detailed view on the learned rules is provided at the project repository.
The reason why approaches like \textsc{RESCAL} and \textsc{ER-MLP} have performed worse than others is probably overfitting.
Embedding methods have shown to achieve excellent results, however no method significantly overperformed the others.
We thus believe that heterogeneity in Linked Data sets is still an open challenge and structure plays a key role to the choice of the algorithm.
Although our MLN framework showed to be more scalable and to be able to provide users with \emph{justifications} for adding triples through the rules it generates, we recognize that this aspect can be further investigated by replacing one or more of its components to decrease the overall runtime.
% reasoning is a powerful resource but not yet efficient.
% Following the examples in other cases (e.g., rule mining, inference), the reasoning task could be limited to the mere transitive closure.
% Avoiding the computation of all consistency and coherence axioms should considerably decrease the overall runtime.

\section{Conclusion} \label{sec:conclusion}
In this paper, we proposed a workflow for probabilistic knowledge discovery as implemented in \mandolin, a framework specifically designed for the Web of Data.
To the best of our knowledge, it is the first complete framework for RDF link prediction based on Markov Logic Networks which features the entire pipeline necessary to achieve this task.
%, including the mining of rules.
We showed that it is able to achieve results beyond the State of the Art for some measures on a well-known link prediction benchmark.
Moreover, it can predict equivalence links across datasets and scale on large graphs.
% Moreover, it can scale on large graphs.
We plan to extend this work in order to refine domain and range in rules and build functionals using OWL rules and evaluate their effectiveness on the predicted links.
% (1) refine domain and range in rules;
% (2) build functionals using OWL rules and evaluate their effectiveness on the predicted links;
% (3) evaluate our approach on triple classification;
% (4) improve the run times with the use of a parallel DBMS and
% (5) a less powerful yet more efficient reasoner.

% \section*{Acknowledgments}
% This work has been supported by the LinkingLOD2 project, funded by the German Research Foundation (DFG) with grant agreement number NG 105/3-2 and URZ Leipzig account 32100441.
% We would like to thank Yang Chen for his precious and kind help.

%% The file named.bst is a bibliography style file for BibTeX 0.99c
\bibliographystyle{aaai}
\bibliography{mandolin,aksw}
\end{document}